\documentclass[aps,prl,twocolumn,showpacs,amsmath,amssymb]{revtex4}

\usepackage{graphicx}

\begin{document}


\title{NMR Observation of Rattling Phonons in the Pyrochlore Superconductor KOs$_2$O$_6$}


\author{M. Yoshida}
\author{K. Arai}
\author{R. Kaido}
\author{M. Takigawa}
\author{S. Yonezawa}
\author{Y. Muraoka}
\author{Z. Hiroi}
\affiliation{Institute for Solid State Physics, University of Tokyo, Kashiwanoha, Kashiwa, Chiba 277-8581, Japan}


\date{\today}

\begin{abstract}
We report nuclear magnetic resonance studies on the $\beta $-pyrochlore oxide
superconductor KOs$_2$O$_6$. The nuclear relaxation at the K sites is entirely caused
by fluctuations of electric field gradient, which we ascribe to highly anharmonic
low frequency oscillation (rattling) of K ions.  A phenomenological analysis shows
a crossover from overdamped to underdamped behavior of the rattling phonons 
with decreasing temperature and its sudden sharpening below the superconducting
transition temperature $T_c$. Suppression of the Hebel-Slichter peak in the relaxation
rate at the O sites below $T_c$ also indicates strong electron-phonon coupling. 
\end{abstract}


\maketitle



The family of new superconductors $A$Os$_2$O$_6$ ($A$ = K, Rb, and Cs) with the
pyrochlore structure exhibits many exotic properties. For instance, the
superconducting transition temperature $T_c$ increases substantially with
decreasing radius of $A$ ions ($T_c$ = 3.3, 6.3, and 9.6 K for $A$ = Cs, Rb, and K,
respectively \cite{Yonezawa1,Yonezawa2,Bruhwiler1,Yonezawa3}) and shows
non-monotonic pressure dependence \cite{Muramatsu}, in contradiction to what would
be expected from the change of density of states alone. Among this family
KOs$_2$O$_6$ is most anomalous.  The electrical resistivity shows strong concave
$T$-dependence down to $T_c$ \cite{Yonezawa1,Hiroi2}, in contrast to the normal
$T^2$-behavior in Rb and Cs compounds \cite{Yonezawa2,Bruhwiler1,Yonezawa3}.  The
$T$-linear coefficient of the specific heat $\gamma $ = 70~mJ/K$^2$mol \cite{Hiroi3,Bruhwiler2} is highly 
enhanced over the value obtained from band calculations 10-11~mJ/K$^2$mol \cite{Kuns1,Saniz}. 

These anomalies imply large density of low energy excitations, which was originally ascribed to magnetic 
frustration inherent to the pyrochlore lattice, a network of corner-shared tetrahedra formed by Os atoms. 
However, the Pauli-like susceptibility \cite{Hiroi3} excludes local moments on the Os sites. How the frustrated lattice 
geometry affects properties of itinerant electrons is still an open issue. On the other hand, there is now 
a growing body of experimental evidence for anharmonic motion of isolated alkaline ions in an oversized 
cage of Os-O network. The specific heat data show existence of low frequency Einstein modes in addition 
to the electronic and the Debye contributions for all members of $A$Os$_2$O$_6$ \cite{Hiroi3,Bruhwiler2,Hiroi1}. 
The X-ray results revealed huge atomic displacement parameter for the K sites in KOs$_2$O$_6$ \cite{Yamaura}. 
Recently a second phase transition at $T_p$ = 7.5 K was observed for high quality single crystals of 
KOs$_2$O$_6$ \cite{Hiroi2}. Since the transition is of first order and persists into the normal state
in high magnetic fields, it is likely a structural transition associated with the K ion dynamics. 
Calculation of effective ionic Hamiltonian also indicates instability of K ions in a highly anharmonic 
potential \cite{Kuns2}, which might be the origin of various anomalies. Oscillations of isolated ions in a 
large space have been discussed also in other materials such as clathrate \cite{Keppens} and 
skutterudite \cite{Paschen} compounds and commonly called ``rattling''.

In this Letter, we report results of nuclear magnetic resonance (NMR) experiments on the $^{39,41}$K 
nuclei (nuclear spin $I$ = 3/2) and $^{17}$O nuclei ($I$ = 5/2) in KOs$_2$O$_6$. 
We found that the nuclear relaxation at the K sites is caused entirely by phonons via the electric 
quadrupole interaction. From a phenomenological analysis we propose that rattling phonons undergo 
a crossover from underdamped to overdamped behavior with increasing temperature. 
Strong electron-phonon coupling is indicated by rapid reduction of the damping below $T_c$. 
On the other hand, the relaxation rate at the O sites is sensitive to electronic excitations. Substantial 
damping of quasiparticles due to rattling phonons near $T_c$ is indicated by suppression of the 
Hebel-Slichter peak at the O sites. A part of the $^{39}$K NMR results has been discussed in terms of spin 
fluctuations in a previous publication~\cite{Arai}, which must now be discarded.

Two powder samples of KOs$_2$O$_6$ were used in the present experiment. The sample A was prepared by heating 
a mixture of KO$_2$ and OsO$_2$ with Ag$_2$O as an oxidizing agency~\cite{Yonezawa1}. 
The sample B was enriched with $^{17}$O by heating KO$_2$ and Os metal in two steps in oxygen atmosphere 
containing 45 \% $^{17}$O. Both samples show identical $T_c$ of 9.6 K. Note that the second transition 
at 7.5~K has never been observed in powder samples.  The nuclear spin-lattice relaxation 
rate $1/T_1$ at the $^{39,41}$K sites was measured by the saturation recovery method. Although the K sites 
have tetrahedral (T$_d$) symmetry, hence should not suffer quadrupole broadening, the $^{39}$K NMR spectra 
consist of a narrow central line (0.3 kHz HWHM at 8.5T) and a slightly broad satellite line (10 kHz HWHM) 
probably due to imperfections. The line shape shows no appreciable temperature ($T$) dependence 
above $T_{c}$.  The entire spectrum was easily saturated by rf comb-pulses, 
resulting in single exponential recovery of the spin echo intensity. The $^{17}$O NMR spectra show a quadrupole 
broadened powder pattern. To determine $1/T_1$, the recovery of the spin-echo intensity of the central line 
as a function of time $t$ after the inversion pulse was fit to the form \cite{Narath}
$M_{eq}-M_{0}\{U\exp(-t/T_1)+V\exp(-6t/T_1)+(1-U-V)\exp(-15t/T_1)\}$ with $U$ and $V$ fixed to the same values 
at all temperatures.  

The $T$-dependences of $1/(T_1T)$ at the $^{39}$K sites is shown in Fig.~1.  The inset of Fig.~1(a) 
displays the isotopic ratio $^{39}T_1/^{41}T_1$ for the sample A. Surprisingly, the isotopic ratio coincides with the 
squared ratio of the nuclear quadrupole moments $(^{41}Q/^{39}Q)^2 =$ 1.48 rather than the nuclear magnetic 
moments $(^{41}\gamma /^{39}\gamma )^2$ = 0.30 in a wide $T$-range 6-100~K. Thus the relaxation is entirely 
caused by fluctuations of electric field gradient (EFG) as opposed to magnetic fluctuations. Since any active electronic 
states should be dominantly $s$-like at the K sites with negligibly small quadrupole coupling to K nuclei, 
it must be phonons that causes the nuclear relaxation. 
\begin{figure}[b]
\includegraphics[width=7.5cm]{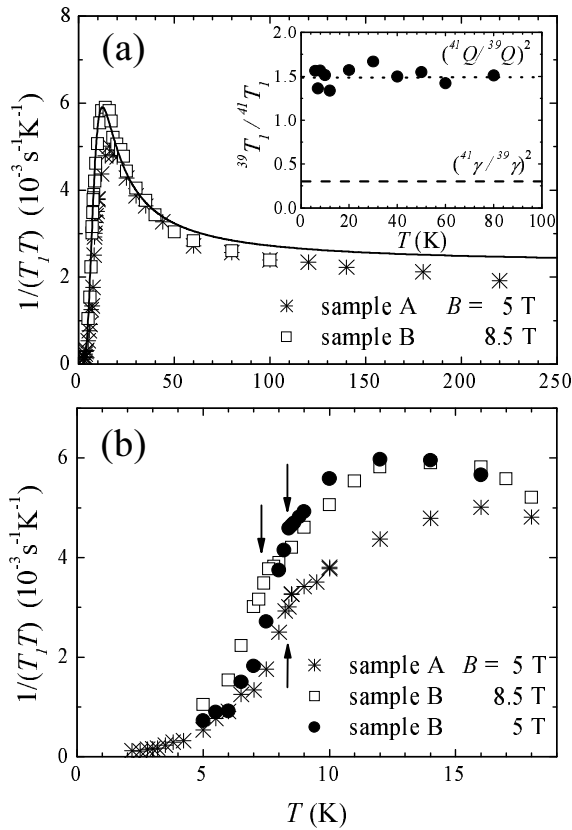}
\caption{\label{fig:fig1} $T$-dependence of $1/(T_1T)$ at the $^{39}$K sites for the two samples
at different magnetic fields. The data for a wide $T$-range are shown in (a).  The solid line shows    
the calculated result of Eq. (5) as described in the text.  The inset shows the isotopic ratio 
of $1/T_1$. The dotted (dashed) line indicates the squared ratio of the nuclear 
quadrupole (magnetic) moments.  The data in the low $T$ region are expanded in (b). The arrows 
show the superconducting $T_c$ at the respective fields.}
\end{figure}

The prominent features of the data in Fig.~1 are: (1) a peak near 13~K (16~K) for the sample B (A), 
(2) approximate constant behavior at high temperatures, (3) rapid decrease at low temperatures, 
(4) a clear kink at the superconducting $T_c$ and sudden decrease below $T_c$. 
The last point indicates strong influence of superconductivity on the phonon dynamics,  
a direct evidence for strong electron-phonon coupling. The two samples show slightly 
different results near the peak temperature. Although the reason is not understood clearly, this may arise from 
different degrees of imperfection or hydration \cite{Hiroi3}.    

For nuclei with spin 3/2, the transition probability $W_q$ between two nuclear levels $|I_z = m \rangle$ and 
$|I_z = m \pm q \rangle$ (q = 1 or 2) due to quadrupole coupling is given by the correlation function of EFG as 
\cite{Abragam} 
\begin{eqnarray}
W_q=\frac{1}{12}\Big(\frac{eQ}{\hbar}\Big)^2\int_{-\infty}^{\infty}\langle [V_{+q}(t),V_{-q}(0)] \rangle e^{iq\omega _Lt}dt, 
\end{eqnarray}
where $\omega _L$ is the nuclear Larmor frequency and $[A, B]$ stands for 
$(AB+BA)/2$. The irreducible components of the 
EFG tensor are defined as $V_{\pm 1} = V_{xz} \pm iV_{yz}$, $V_{\pm 2} = (V_{xx} - V_{yy} \pm 2iV_{xy})/2$, where 
$V_{xy} = \partial^2 V/\partial x\partial y$ etc. is the second derivative of the electrostatic potential at the nucleus.  For simplicity we assume spherically symmetric 
EFG fluctuations.  Then $W_1 = W_2 \equiv W$ and the relaxation rate is given as $1/T_1 = 2W$. 

In diamagnetic insulators phonons are usually the dominant source of relaxation for quadrupolar nuclei. 
The ionic motion modulates the EFG, 
\begin{eqnarray}
V_{\pm q}=V_{\pm q,0} + \frac{\partial V_{\pm q}}{\partial u}u + 
\frac{1}{2}\frac{\partial ^2V_{\pm q}}{\partial u^2}u^2+\cdot \cdot \cdot ,
\end{eqnarray}
where $u$ is the ionic displacement from the equilibrium position expressed as a linear combination of the phonon 
creation ($a^\dagger $) and annihilation ($a$) operators. The second term inserted into Eq.~(1) leads to 
the \textit{direct process} \cite{Abragam}, which is expressed by the one phonon correlation 
function or the imaginary part of the susceptibility, 
\begin{eqnarray}
W_q \propto k_BT\frac{\mathrm{Im}\chi (q\omega _L)}{q\hbar \omega _L},
\end{eqnarray}
provided $q \hbar \omega_{L} \ll k_{B}T$. For harmonic phonons with infinite life time, the contribution from the direct process 
is negligible since $\omega _L$ ($\sim$ 10 MHz) is many orders of magnitude smaller than the typical phonon frequency and the 
phonon density of states at $\omega _L$ is practically zero. 

On the other hand, the third term in Eq.~(2) leads to the \textit{two phonon Raman process}, in which a nuclear transition occurs by 
absorbing one phonon and emitting another phonon \cite{Abragam}. Usually this is by far the dominant process expressed 
as $1/T_1 \propto \int \{ \rho (\omega) \}^{2} n(\omega) (n(\omega) + 1) |A(\omega)|^{2} d\omega$, where $n(\omega)$ is the 
Bose factor, $\rho (\omega)$ is the phonon density of states, and $|A(\omega)|$ is the transition matrix element. One can 
see $1/T_1 \propto T^2$ at high temperatures where $n(\omega) \sim T/\omega $. At low temperatures, $1/T_1$ 
decreases monotonically and approaches either to $T^7$-dependence for acoustic phonons or activated behavior for optical 
phonons \cite{Abragam}. Our results in Fig.~1 are, however, in strong contradiction with such behavior.

Since dominance of phonons for nuclear relaxation is extremely rare in metals, we suppose that rattling makes a special case 
for KOs$_2$O$_6$. The distinct features of rattling are its low energy scale and strong anharmonicity, 
which should lead to damping. We expect that damping-induced broadening of the phonon spectrum peaked at a very 
low frequency, will bring substantial spectral weight at zero frequency so that the direct process (Eq.~3) becomes important. 
The peak in $1/(T_1T)$ is then naturally explained as follows. Let us assume a low frequency 
Einstein mode at $\omega _0$. At low temperatures, the spectrum $\mathrm{Im}\chi(\omega )/\omega $ is sharp, therefore, the weight 
at $\omega _L \approx 0$ should be small.  As the spectral width increases with increasing temperature, the weight at $\omega = 0$ 
will grow and take a maximum when the width becomes comparable to $\omega _0$. Further broadening, however, will spread 
the spectrum over higher frequencies and reduce the weight at $\omega = 0$. (See the inset of Fig.~2.) Thus the peak in $1/(T_1T)$ 
can be reproduced if the rattling undergoes a crossover from underdamped to overdamped behavior with increasing temperature. 
The contribution from the Raman process should be also influenced by damping \cite{Rubinstein}.  However, we could not reproduce the peak in $1/(T_1T)$ by the
Raman process without fine tuning the $T$-dependence of the damping.  We consider
that the Raman process is unimportant. The details will be discussed elsewhere \cite{Yoshida}.
\begin{figure}[b]
\includegraphics[width=7.5cm]{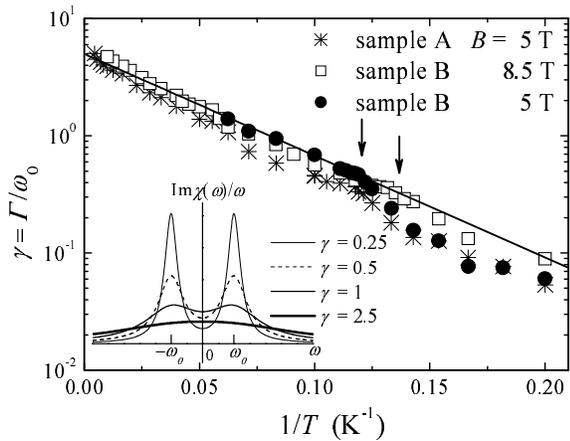}
\caption{\label{fig:fig2} The normalized damping $ \gamma = \Gamma /\omega _0$ is plotted against $1/T$. 
The arrows correspond to $T_c$ at 5~T and 8.5~T.  The inset shows the phonon line shape Im$\chi (\omega )/\omega$ 
calculated from Eq.~(4) for several values of $\gamma$.}
\end{figure}

In order to quantify the above picture, we employ a suitable phenomenological model describing strongly damped oscillators. 
A commonly used expression is the damped harmonic oscillator model, 
$\mathrm{Im}\chi (\omega )/\omega = \Gamma /\{(\omega _0^2 - \omega ^2)^2 + \Gamma ^2\omega ^2\}$, which leads to 
$1/(T_1T)\propto \Gamma /\omega _0^4$ by taking the limit $\omega _L \rightarrow 0$ in Eq. (3). However, this model is unable 
to reproduce the observed peak for any monotonic change of the damping $\Gamma $. In the damped harmonic oscillator model, 
only the friction or the momentum relaxation is considered as the sources of damping. It was argued, however, that in 
the collision dominated regime one must also consider the jump of the spatial coordinate or the amplitude relaxation \cite{Silverman,Takagi}.  
One then obtains the van Vleck-Weisskopf formula \cite{VanVleck}, which should be suitable for 
strongly damped anharmonic phonons, 
\begin{eqnarray}
\frac{\mathrm{Im}\chi (\omega )}{\omega } = \frac{\Gamma /\omega _0 }{(\omega + \omega _0)^2+\Gamma ^2} + 
\frac{\Gamma /\omega _0 }{(\omega - \omega _0)^2+\Gamma ^2}.
\end{eqnarray}
From Eqs.~(4) and (3) we obtain in the limit $\omega _L \rightarrow 0$,  
\begin{eqnarray}
\frac{1}{T_1T} \propto \frac{1}{\omega _0^2}\frac{\gamma }{1+\gamma ^2},
\end{eqnarray}
where $\gamma = \Gamma /\omega _0$. For a fixed value of $\omega _0$, this formula has a maximum at $\gamma = 1$, which 
corresponds to the observed peak.  By identifying the $T$-ranges below or above the peak temperature with 
the underdamped ($\gamma \leq 1$) or the overdamped ($\gamma \geq 1$) regions and assuming that $\omega _0$ is constant, 
we extracted the T-dependence of $\gamma $ from the experimental data of $1/(T_1T)$ as shown in Fig. 2. The result above 
$T_c$ is well represented by an activation law $\gamma  = \gamma _0$exp$(-E/k_BT)$ with $\gamma _0 = 5$ and $E =$ 20 K (the solid line). 
Thus our analysis predicts extremely overdamped behavior at high temperatures. Note that the phonon frequency $\omega _0$ cannot be 
determined from our analysis. The solid line in Fig. 1 shows the $T$-dependence of $1/(T_1T)$ calculated from Eq. (5) and 
this activation law. A possible interpretation of the activated behavior is that $\Gamma $ represents the hopping frequency 
of the K ions among different metastable positions in highly anharmonic potential \cite{Kuns2} and the activation energy 
corresponds to the potential barrier between them. 
\begin{figure}[b]
\includegraphics[width=7.5cm]{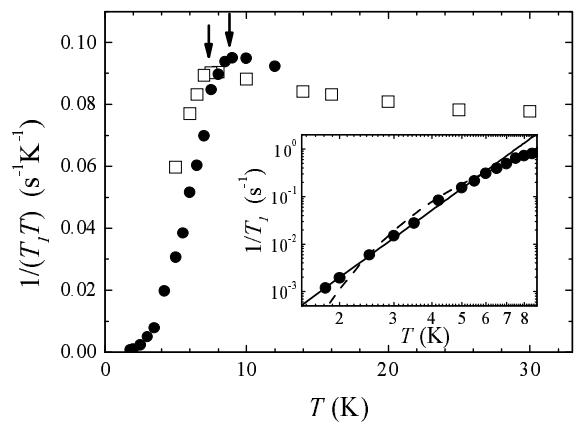}
\caption{\label{fig:fig3} $T$-dependence of $1/(T_1T)$ at the $^{17}$O 
sites at 2 T (solid circles) and 8.5 T (open squares). The arrows indicate $T_c$ at the respective fields. 
The inset shows the fitting to a power-law $1/T_1\propto T^\alpha $ with $\alpha  = 4.7$ (solid line)
and an activation law $1/T_1\propto \exp(\Delta /T)$ with $\Delta  = 17$ K 
(dashed line) for the data at 2 T below $T_c$.}
\end{figure}

We have also measured the relaxation rate for $^{85}$Rb and $^{87}$Rb 
nuclei in RbOs$_2$O$_6$ and separated the phononic and magnetic contributions \cite{Yoshida}. The phonon 
contribution to $1/(T_1T)$ in RbOs$_2$O$_6$ does not show a peak, indicating underdamped behavior in the whole $T$-range 
below the room temperature. The $T$-dependence of $\gamma$ in RbOs$_2$O$_6$ obtained from similar analysis 
is also compatible with an activation law with $E$=12 - 20~K, although the accuracy is limited.  

A remarkable feature of the results in Fig. 2 is the sudden reduction of $\gamma $ below the superconducting $T_c$. 
This indicates that the damping of rattling phonons near $T_c$ is primarily caused by electron-phonon interaction 
and the phonon life time is enhanced by opening of the superconducting gap. Therefore, $\omega_{0}$ should be 
smaller than 2$\Delta \sim 50$~K \cite{Hiroi3}, where $\Delta$ is the superconducting gap. 

We now discuss the results at the O sites. Figure 3 shows the $T$-dependence of $1/(T_1T)$ at the O sites at the fields 
of 2 T and 8.5 T. Unlike the result for the K sites, $1/(T_1T)$ is only weakly $T$-dependent above $T_c$, approximately 
consistent with the Korringa relation. Thus the spin dynamics of conduction electrons are probed at the O sites. 
This is consistent with the Os-5$d(t_{2g})$ and O-2$p$ hybridized nature of the conduction band \cite{Kuns2,Saniz}.  
However, the weak $T$-dependence of $1/(T_1T)$ in the normal state is still anomalous and suggests development of 
modest spin correlation.  

Below $T_{c}$, it is known that $1/(T_1T)$ shows a peak near $T=0.9T_{c}$ (the Hebel-Slichter peak \cite{Hebel})
in weak coupling superconductors with an isotropic gap.  This is not the case for the data in Fig. 3; $1/(T_1T)$ decreases
gradually below $T_{c}$, indicating that the Hebel-Slichter peak is strongly suppressed.  
At lower temperatures,  $T$-dependence of $1/T_1$ can be fit either by a power law $1/T_1 \propto T^\alpha $ 
with $\alpha \sim 5$ or by an activation law $1/T_1 \propto \mathrm{exp}(-\Delta /T)$ with $\Delta = 17 K$. 
Since the upper critical field $H_{c2}$ is about 30 T at $T = 0$ \cite{Ohmichi,Shibauchi}, pair breaking effects 
of magnetic field should not be strong at 2 T \cite{Pennington}.

The behavior at low temperatures rules out highly anisotropic gap structure such as line nodes, which leads to $T^3$ dependence 
of $1/T_1$. This is consistent with the field independent thermal conductivity showing an isotropic gap \cite{Kasahara}. 
Therefore, suppression of the Hebel-Slichter peak cannot be ascribed to a highly anisotropic gap. It appears more likely 
due to inelastic scattering of quasiparticles by phonons, which broadens the superconducting density of states as was 
proposed for the case of TlMo$_6$Se$_{7.5}$ \cite{Kitaoka}.   
This has been also predicted by numerical calculations based on the Eliashberg theory \cite{Allen,Akis}. 
The calculation by Akis {\it et al}. \cite{Akis}, for example, shows that the Hebel-Slichter peak disappears when 
$k_BT_c/\hbar \omega _{ln}$ is larger than $0.2 \sim 0.3$, where $\omega _{ln}$ is the Allen-Dynes parameter representing 
the typical phonon frequency. This is compatible with our previous conclusion that the frequency of the rattling phonon is 
smaller than $2\Delta \sim 5k_BT_c$. Whether the rattling works as a pairing mechanism or acts as a pair breaker is still 
an open question. 

In conclusion, our analysis of the $1/(T_1T)$ data at the K sites indicates a crossover of rattling from the low $T$ underdamped to 
high $T$ overdamped behavior near 13 - 16~K.  The rapid decrease of $1/(T_1T)$ at the K sites below $T_c$ indicates 
strong electron-phonon coupling and enhancement of the lifetime of rattling phonons.  Hence its frequency must be smaller 
than $2\Delta $. The results at the O sites rule out strongly anisotropic gap. The absence of the Hebel-Slichter peak at the O 
sites is also compatible with strong electron-phonon coupling.

We thank T. Kato, Y. Takada, H. Maebashi K. Ishida, H. Harima, H. Fukuyama, Y. Matsuda, and 
S. Tajima for valuable discussions.  This work was supported by a Grant-in-Aid for Scientific 
Research on Priority Areas 
(No. 16076204 ``Invention of Anomalous Quantum Materials'') from the MEXT Japan.

\end{document}